# Hardware Accelerated Protein Inference Framework

S.M. Vidanagamachchi, S.D. Dewasurendra, *Member, IEEE* and R.G. Ragel, *Member, IEEE*

*Abstract*- Protein inference plays a vital role in the proteomics study. Two major approaches could be used to handle the problem of protein inference; top-down and bottom-up. This paper presents a framework for protein inference, which uses hardware accelerated protein inference framework for handling the most important step in a bottom-up approach, viz. peptide identification during the assembling process. In our framework, identified peptides and their probabilities are used to predict the most suitable reference protein cluster for a given input amino acid sequence with the probability of identified peptides. The framework is developed on an FPGA where hardware software co-design techniques are used to accelerate the computationally intensive parts of the protein inference process. In the paper we have measured, compared and reported the time taken for the protein inference process in our framework against a pure software implementation.

## I. INTRODUCTION

Protein inference is a critical step in the proteomics study, which involves assembling peptides identified from Tandem Mass Spectrometry (MS/MS experiment) into a set of proteins [4]. Protein inference can be performed either as top-down and bottom-up approaches. The top-down approach involves intact proteins and it identifies proteins from currently available masses/mass spectrum databases by performing a database search. Here intact protein data is converted into mass spectrum data, which can be processed by digital computers. The bottom-up approach (shotgun proteomics) involves digestion of a complex protein sample into peptides and then inferring the proteins from the analysed peptide data and their characteristics [1]. Fig. 1 shows the steps in the process of a typical bottom-up protein inference. The protein sample is first digested into peptides using some enzyme such as trypsin. Then the mass spectrum data of the peptide set is generated. Next a database search should be performed to identify the peptides. Finally identified peptides are assembled to infer the proteins.

The bottom-up approach is the standard, commonly used strategy which has also been proven as a successful and robust approach [1]. However, issues such as peptide degeneracy still remain unsolved. Degenerate peptides are the peptides shared by multiple proteins that lead to ambiguities in the identification process. Further the protein inference is difficult due to not having a complete relation between peptides identified and parent proteins that may be present.

Top-down and bottom-up approaches have their advantages and disadvantages. No digestion step or peptide identification step is required in top-down approach and it makes it possible to identify different isoforms or post-translational modifications of proteins which could not be achieved in bottom-up approach (due to hard characterization and quantifying of intact proteins from a complex sample of cleaved peptides [3] and due to degenerate peptides). However top-down approach is not well suited to study unknown proteins such as unsuspected modifications or sequence variants (the database used in top-down approach normally has one spectrum of the protein and modifications in the protein will prevent the chance of getting a match). Therefore, one needs to have one's own database consisting of all mutants and splice variants of each protein since standard sequence databases might not represent the intact protein by default.

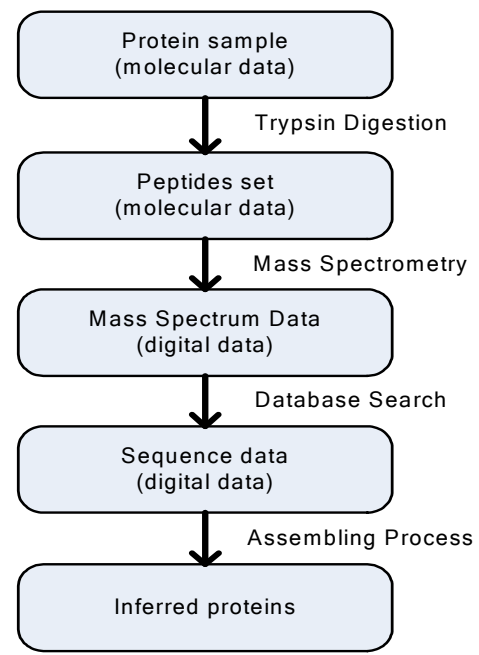

Fig. 1. Steps in the bottom-up approach

In the bottom-up approach, the advantage is that when some peptides are lost due to proteins that have mutants or splice variants, some matches of peptides could still be obtained when inferring the proteins.

In this paper, we discuss our protein inference framework that uses the bottom-up approach. The main target of our framework is to accelerate the process of protein inference by developing it using a hardware software co-design process in an FPGA. The key idea is in the calibration process of the framework where it is used, first to identify the occurrence of peptides in a given set of hypothetical (and unknown) sequences which are then used to compute probabilities that can later be used in inferring unknown proteins submitted to the framework, also with a probability. Hence, once properly set up, given a set of peptide sequences from a complex protein sample at the input, the framework can infer the protein. For our prototype system, we have mapped the maximum possible number of peptides to the Altera Cyclone II FPGA. However, we have to deal with degenerate or shared peptides to come up with a more accurate system, details of which are discussed later in the paper.

We input a homologous unknown protein sequence to the framework and then it identifies the peptides in it. This result is then used to compute the probability of occurrence of each such peptide in the input protein. Here we have used absolute probability and the sample size of 10 for our experiments. The next task is to use this information in bottom-up protein inference work chain: our contribution here is in the last step of the workflow, viz., starting with sequence data, inferring the protein. We do this by first identifying the peptides in these sequences and then using the probabilities computed above using our framework to infer the protein: the reference protein which gives the highest total probability sum for containing the set of identified peptides will get selected.

The most important task of our framework is to accelerate the matching process of the input protein or a set of peptide sequences against known peptides. This task is performed on Aho-Corasick automata for several peptide clusters which are arranged as tiles in the tile architecture of FPGA. Each cluster consists of several peptides which represents UniProt reference proteins. Hardware software co-design was used for accelerating the peptide matching process, in which C macros are used to access the hardware logic of FPGA through Avalon Memory Mapped Interface. Interfacing the input and output, calculating and matching probabilities of the peptides and inferring proteins with maximum probabilities are performed in the software design. The entire process of hardware software co-design is presented later in the paper.

The paper is organised as follows: we discuss the related work on protein inference in Section II. Section III describes the methodology we have used in the experiments and the obtained results. Conclusions are discussed in Section IV and future work is included in Section V.

## II. RELATED WORK

A heuristic algorithm has been proposed by Pedro et al. for solving peptide degeneracy problem by using peptide detectability [2]. They have dealt with top-down approach. John et al. have discussed the techniques used in top-down proteomics and how it could be used in clinical research [3].

Several strategies of inferring proteins from identified peptides and the challenges in this process were discussed by Yong et al. in their review [4]. According to the review, strategies are categorized as, rule based strategies, combinatorial optimization algorithms and probabilistic inference algorithms. Homology based protein function prediction has been performed by Hamp et al. [5]. A probabilistic protein inference technique has been proposed by Li and Radivojac for characterizing proteomes [6]. Bottom-up protein inference problem was addressed by Alexey et al. and they have analyzed shotgun proteomics considering degenerate peptides [7]. In our approach we have also followed bottom-up approach and we have used a probability based method for the inference. It differs from the above mentioned methods because it does not use probabilities of proteins appearing in the mixture or it does not use any optimization algorithm. It only considers the collective (total) absolute probability of appearance of a set of peptides in a particular protein for inference of that protein.

There are no hardware implementations developed for inferring proteins reported in literature, whereas some other hardware accelerated methods have been developed in the past for peptide and protein identification. Yoginder et al. have developed a hardware accelerated system with FPGAs for identifying peptides in chromosome 1 of the human genome [8]. Area optimization strategy of FPGA has been introduced by Vidanagamachchi et al. for accelerating peptide identification [9].

## III. METHODOLOGY

To follow the bottom-up approach in our method we have used the ideas of sequence clustering of peptides in the last two stages of bottom-up approach as shown in Fig. 1. Our approach could be used in the assembling process to infer proteins. Sequence clustering is a method of grouping the biological sequences that are somehow related. These sequences could be genes, transcriptomes or proteins. In general proteins are clustered into homologous families. There exist several methods of protein clustering and many of them use several statistical models for clustering such as Bayesian approach, Q-FISH, etc. [2]. Several software packages have been developed in sequence clustering such as USEARCH[1] and UICluster[2]. These packages also use some models for clustering in the algorithm. Protein identification based on identified peptides is a difficult task and it is tough to develop an effective model. Use of proteotypic peptides or proteotypic peptide libraries is one approach to accomplish this task. Proteotypic peptides are peptides that are most likely to be identified by mass spectrum based techniques.

Clustering of peptides in order to group degenerate peptides was performed in IsoformResolver software by Meyer et al., which has a mapping of proteins and peptides from a peptide centric database-a database of map files which maps peptides to corresponding proteins [11]. In our experiment, we have performed the clustering based on UniProt reference clusters[3], which consist of reference protein sequences for several homologous proteins. We have a mapping of reference clusters into sets of peptides. Here a protein sequence is split into its corresponding peptides set, which are then arranged to represent clusters. Therefore, we could use one reference sequence for homologous proteins which are derived from a common ancestor. This methodology has not been used in the past to infer proteins.

---

[1] http://www.drive5.com/usearch/
[2] https://genome.uiowa.edu/pubsoft/clustering/
[3] http://www.uniprot.org/

Creation of protein clusters is based on peptides from homologous proteins. Hence, our system should include degenerate peptides which we hope to consider in our future work. Degenerate peptides could cause reduction in identification probability of peptides in some special cases. The concept of peptide detectability is one way of handling degenerate peptides. It is defined as the probability of identifying peptides in an MS/MS experiment. MSBayesPro[4] (Bayesian Protein Inference algorithm) is one algorithm that is developed using the ideas of peptide detectability. In this paper, we concentrate on searching the maximum number of peptides that can be mapped to a protein using their identification probabilities. Further this could be used for homology based protein function prediction.

Bit-split version of Aho Corasick algorithm is used to implement our clustering framework in hardware. Bit split version of Aho Corasick algorithm is known to be faster than non-bit split Aho Corasick algorithm [8]. In the first stage of the algorithm it constructs a finite state machine with the given set of keywords and then it locates the keywords in an input text string. We have used this for identifying peptides in unknown/predicted protein sequences in a hypothetical database (HypoDB[5], HAPPY [10]) and then find the absolute probabilities of those peptides appearing in the protein reference sequences. Then we compute the largest total probability of appearance, which we use as our criterion for guessing the protein. Since, the Aho Corasick algorithm uses several Finite State Machines (FSM), the storage space of FPGA for pattern mapping is the major limitation here. Therefore we have mapped the maximum number of peptides in the selected set of clusters and they are mapped into 97% of logic elements in our FPGA. Since we have used peptides of reference protein clusters, one reference protein is used to represent many proteins and therefore many peptides in different organisms.

Avalon Memory Mapped interface is used to make the communication between Nios II processor and user defined custom logic (cluster based FSM logic). Fig. 2 shows how the communication is performed between an FSM and the Nios II processor via Avalon MM interface. In our system we have 5x20 FSMs and all of them work in a similar way. In the logic automata created, we are using 315 peptides and they are mapped to the area of FPGA to represent 13 clusters in 20 tiles (Table I). This type of hardware implementation has not been done in the past.

The implementation of the prototype is completed on a Cyclone II FPGA chip by Altera. According to the Avalon Interface Specification[6] we could add our custom logic into SOPC builder[7] separately. Then it acts as a separate custom component through Avalon MM interface. There are two types of Avalon ports: master and slave ports. Master ports initiate transfers and slave ports respond to transfer requests. They provide common interfacing signals such as address, **clk**, **readdata**, **writedata**, **write** (a write enable signal), **read** (a read enable signal), **chipselect**, and **waitrequest** to support simple transfers. They further support complex transfers such as latent and streaming transfers.

SOPC builder consists of the following functional blocks: Peripheral task logic that performs functional operations of the component, Register file that provides memory elements for input to and output from task logic, Avalon interface that provides a standard address mapped interface to the register file, Software driver functions that provide software application interface to the component (specially for read data from and write data to the component).

By using the Avalon Memory Mapped slave interface, we have connected our custom component logic to the system interconnect fabric which is responsible for connecting components together in the SOPC builder. Moreover the systems interconnect fabric supports transactions performed by Avalon masters and slaves for Avalon Memory Mapped components. Then we run the custom logic in Nios II processor (a soft-core by Altera) which runs the prototype protein inference framework that consists of 319 peptides extracted from 13 protein clusters. In hardware, we have mapped these peptides into a tile architecture where each tile contains a maximum of 20 peptides. Our prototype has 20 tiles (Fig. 3). Table 1 gives details of corresponding tiles, clusters and number of peptides in our framework.

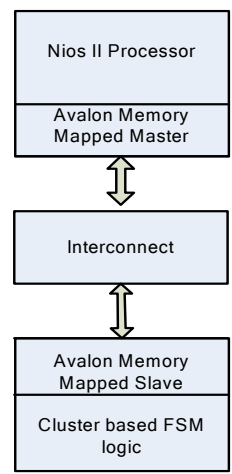

Fig. 2. Add custom component to Nios II processor

The entire protein inference framework that we have developed is shown in Fig. 4, and where the first three steps involve splitting and clustering of peptides in order to map them to the tiles: each of them representing a reference protein. After making the tile architecture from the clustered peptide set we could match our hypothetical amino acid sequences against the hardware. In the final step absolute probabilities of identified peptides are calculated, using which protein/s can be inferred.

---

[4] http://darwin.informatics.indiana.edu/yonli/proteininfer/
[5] http://www.bioclues.org/hypo/index.php
[6] http://www.altera.com/literature/manual/mnl_avalon_spec.pdf
[7] http://www.altera.com/literature/lit-sop.jsp

Interfacing the input and output is done through Nios II IDE. Calculating matching probabilities of the peptides and inferring proteins with maximum probabilities are performed in the software design. User should give the input (in case of multiple proteins) to the system as comma separated input. Our hardware component consists of several tiles, each representing a protein reference cluster. The tiles can execute in parallel during the matching process. Each tile consists of a Finite State Machine built according to the Aho Corasick algorithm for mapping peptides into tiles.

## IV. RESULTS AND DISCUSSION

We have tested the corresponding reference protein sequences to validate of our system. The average times of computation are given in Fig. 5. and Fig. 6. for two types of design: software only and hardware software co-design, respectively. Here total time of hardware software co-design includes the times taken for writing inputs to custom logic, matching the peptides in the protein for identification of peptides, reading the result back and calculation of absolute probability of being in the identified protein. In software only implementation total time is the time taken for matching process and the time taken for inference the protein by probability calculation. Fig. 7. shows the ratios of speedups of hardware software co-design compared to software only implementation in a Nios II processor. This speed up is approximately 19 in matching process and 18 in total process. We will be extending this system for further analysing peptides considering parsimony principle, etc.

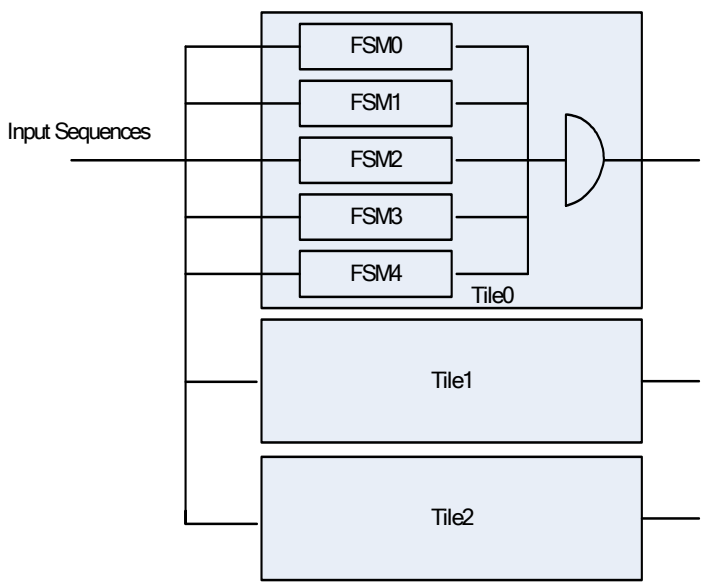

Fig. 3. Tile Architecture

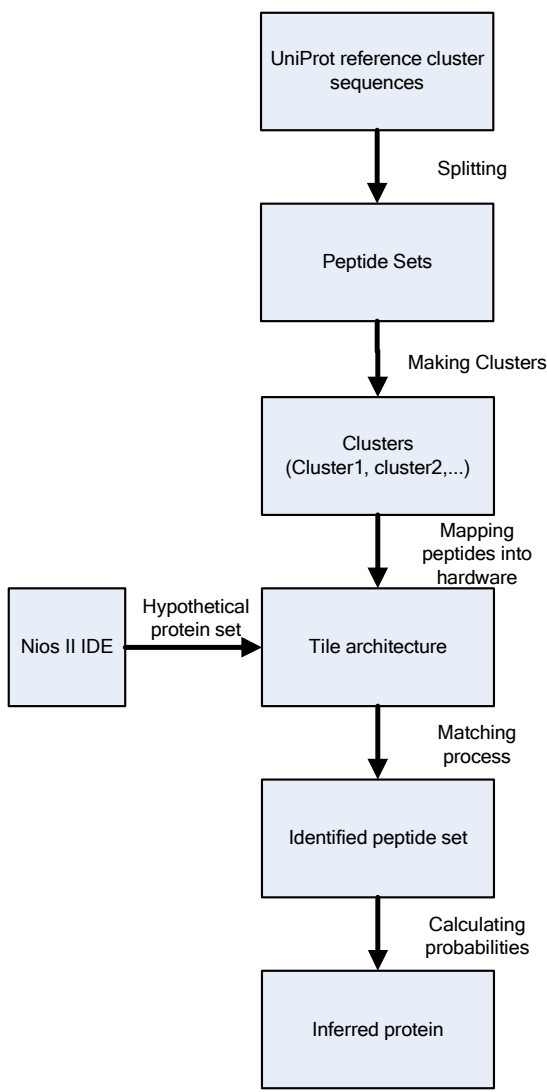

Fig. 4. Protein Inference Framework

TABLE I
MAPPING PROTEIN REFERENCE CLUSTERS TO TILES

| Tile No. | Cluster No. | No. of peptides |
|---|---|---|
| 0 | 1 | 13 |
| 1 | 1 | 14 |
| 2 | 2 | 09 |
| 3 | 3 | 20 |
| 4 | 3 | 19 |
| 5 | 4 | 18 |
| 6 | 5 | 12 |
| 7 | 6 | 18 |
| 8 | 7 | 20 |
| 9 | 7 | 20 |
| 10 | 7 | 18 |
| 11 | 8 | 20 |
| 12 | 8 | 20 |
| 13 | 8 | 19 |
| 14 | 9 | 11 |
| 15 | 10 | 14 |
| 16 | 11 | 5 |
| 17 | 12 | 17 |
| 18 | 13 | 16 |
| 19 | 13 | 16 |

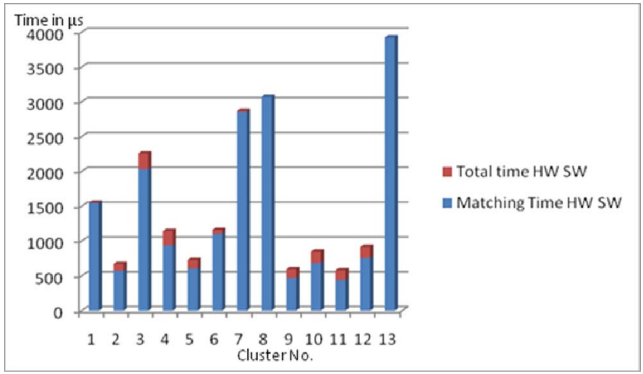

Fig. 5. Total time and matching time taken in HW SW co-design

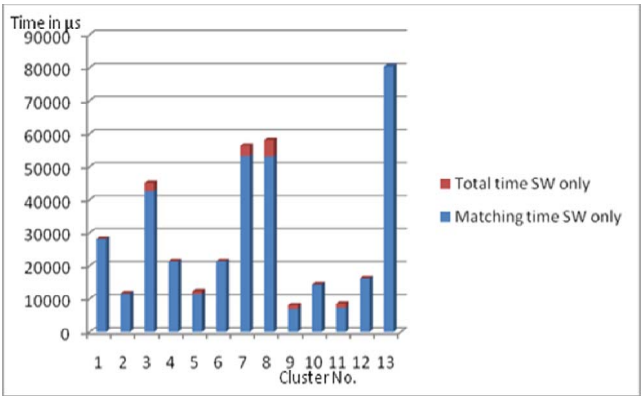

Fig. 6. Total time and matching time taken in software only design

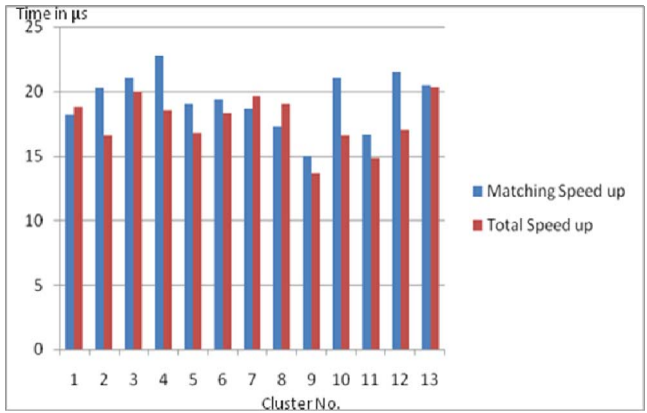

Fig. 7. Ratios of HW SW co-design with SW only implementation

## V. CONCLUSION

In this prototype framework we present a methodology of identifying proteins, which could be extended in a suitable manner using an FPGA with higher number of logic elements in order to use it as a real system. Currently it can handle only a smaller number of peptides compared to realistic numbers of most proteins. As the result we do not get an exact member sequence in the cluster, instead we get the cluster reference sequence, which represent several member sequences as the results.

## VI. FUTURE WORK

We planned to extend this protein inference framework by considering other factors degenerate peptides, homology based function prediction and develop a more accurate framework. Further we wish to design and implement hardware optimization methods for this framework using Aho-Corasick algorithm and a homology based protein function prediction as our future work.